\documentstyle[11pt,mrs2001,epsfig]{article}
\begin{document}
\title{GALAXY SPECTRAL TYPES AND THEIR APPLICATIONS IN THE 2dFGRS}

\author{DARREN S. MADGWICK \& OFER LAHAV}
\affil{Institute of Astronomy, University of Cambridge, Madingley
Road, Cambridge CB3 0HA, UK}

\begin{abstract}
The 2dF Galaxy Redshift Survey (2dFGRS) has already 
measured over 200,000 redshifts of nearby (median redshift $z\sim0.1$) 
galaxies.  This is
the single largest set of galaxy spectra ever collected.
It allows us to subdivide the survey into subsets according to the 
galaxy intrinsic properties. 
We outline here a method (based on Principal Component Analysis) 
for spectrally classifying these galaxies in a 
robust and independent manner.  In so doing we develop a continuous measure
of spectral type, $\eta$, which reflects the actual distribution of
spectra in the observed galaxy population.  We  demonstrate the usefulness
of this classification by estimating luminosity
functions and clustering per spectral type.
\end{abstract}

\section{Introduction}
The 2dF Galaxy Redshift Survey (2dFGRS) is  an ambitious project with the
aim of mapping the galaxy distribution of the local universe more accurately
than ever before~\cite{colless}.   
It aims to acquire a complete sample of $\sim$250,000 galaxy spectra, 
down to an
extinction corrected magnitude limit of $b_J<19.45$.  The survey is now
approaching completion with over 200,000 galaxy redshifts having already
been measured (as of October 2001), 
and is already starting to yield significant scientific 
results (e.g \cite{norberg}, \cite{peacock}, \cite{percival}, 
\cite{madgwick}). 

Apart from the main science goals of quantifying the large-scale structure of
the Universe, one of the most significant contributions of the 2dFGRS
is to our understanding of the galaxy population itself.  Having a data set of 
250,000 galaxy spectra allows us to test the validity of galaxy 
formation and evolution scenarios with unprecedented accuracy.  However, 
the shear size of the data set presents us with its own unique problems. 
Clearly in order to make the data set more `digestible' some form of data
compression is necessary, whether that be in the form of equivalent width 
measurements, morphological types, colours or some other compression.  
These quantities can be compared with theoretical predictions and
simulations, and hence can set constraints on scenarios for 
galaxy formation and biasing.
The approach we outline here
is the development of a spectral classification derived from the 
distribution of the galaxy spectra themselves, using a 
Principal Component Analysis (PCA).  Using the PCA as the
basis of our classification has distinct
advantages over other methods.  Most notably the data are allowed to `speak for
themselves' rather than having ad-hoc classifications imposed
on them.  In addition, unlike equivalent widths or colours the PCA uses all 
the information
contained in any given spectrum.  The advantage of this is not only the fact 
that the classification is more informative but also that it
is more useful at lower signal-to-noise ratios where individual spectral
channels become noisy.
Other data-compression and classification methods have been 
considered recently in other studies (e.g. \cite{slonim}, \cite{heavens}, \cite{lahav}). 

In these proceedings we briefly outline the development of our spectral
classification and demonstrate several applications to which it has been put.
For more details see \cite{madgwick}.

\section{Spectral Classification}

\subsection{Principal Component Analysis}
 
Principal component analysis (PCA) is a well established statistical technique 
which has proved
very useful in dealing with highly dimensional data sets.  In the 
particular case of galaxy spectra we are typically presented with 
approximately 
1000 spectral channels per galaxy, however when used in applications this is
 usually compressed down to just a
few numbers, either by integrating over small line features - yielding 
equivalent
widths - or over wide colour filters.  The key advantage of using PCA in 
our data compression is 
that it allows us to make use of all the information contained in the spectrum
in a statistically unbiased way, i.e. without the use of such ad hoc filters.

In order to perform the PCA on our galaxy spectra we first construct a 
representative volume limited sample of the galaxies.  When we apply
the PCA to  
this sample it constructs an orthogonal set of components (eigenspectra)
which span the wavelength space occupied by the galaxy spectra.  These 
components have been specifically chosen by the PCA in such a way that as much
information (variance) is contained in the first eigenspectrum as possible, 
and that the amount of the remaining information in all the subsequent 
eigenspectra is likewise maximised.  Therefore, if the information contained
in the first $n$ eigenspectra is found to be significantly greater than that
in the remaining eigenspectra we can significantly compress the data set
by swapping each galaxy spectrum (described by 1000 channels) with its 
projections onto just those $n$ eigenspectra.

In the case of galaxy spectra we find that approximately two thirds of the 
total variance (including the noise) in the spectra can be represented in terms of only the first
two projections ($pc_1$, $pc_2$).  So, at least to a first approximation, galaxy
spectra can be thought of as a two dimensional sequence in terms of these
two projections.

\begin{figure}
 \plotone{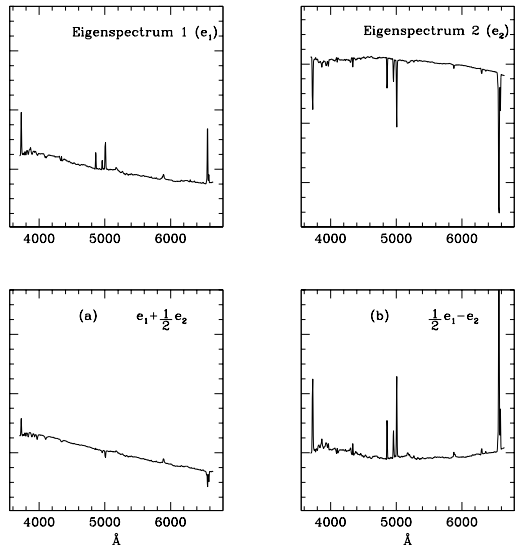}{8cm}
 \caption{The first two principal components identified from the 2dF
 galaxy spectra and the linear combinations which either (a) isolate
 the continuum slope or (b) the line features.}
 \label{evec}
\end{figure}

In Fig.~\ref{evec} we show these first two eigenspectra.  It can be seen from 
this figure that whilst the first eigenspectrum contains both information from
the continuum and lines, the second is dominated by the latter.
Because of this it is possible to take two simple
linear combinations which isolate either the continuum or the 
emission/absorption
line features.  In effect what we are doing when we utilise these linear 
combinations is rotating the axes defined by
the PCA to make the interpretation of the components more
straightforward. In so doing we can see that a parameterisation in
terms of $pc_1$ and $pc_2$ is essentially equivalent to a two
dimensional sequence 
in colour (continuum slope) and the average emission/absorption line
strength.

\begin{table}
 \begin{center}
 \caption{The relative importance (measured as variance) of the first
 8 principal components.}
 \begin{tabular}{@{}cccc}
   \hline
   Component & Variance (\%) & Component & Variance (\%)\\
   \hline
     1 & 51 & 5 & 1.7 \\
     2 & 15 & 6 & 1.3 \\
     3 & 3.4 & 7 & 0.99 \\
     4 & 2.4 & 8 & 0.84 \\
   \hline
 \end{tabular}
 \end{center}
 \label{tabpc}
\end{table}

\subsection{$\eta$ Parameterisation}

The 2dF instrument~\cite{lewis} was designed to measure large numbers of 
redshifts in as short an observing time as possible.  
%In so doing it has made 
%very ambitious projects such as the 2dFGRS possible. 
However, in order to 
optimise the number of
redshifts that can be measured in a given period of time, compromises
have had to be made with respect to the spectral quality of the
observations.  Therefore if one wishes to characterise the observed
galaxy population in terms of their spectral properties care must be
taken in order to ensure that these properties are robust to the
instrumental uncertainties.

The 2dF instrument makes use of up to 400 optical fibres with a
diameter of 140$\mu$m (corresponding to $\sim2.0''$ on the sky,
depending on 
plate position).  The quality and representativeness of the observed
spectra can be compromised in several ways and a full list is
presented in previous work~\cite{madgwick}.  The net effect is that
the uncertainties introduced into the fibre-spectra
predominantly effect the calibration of the continuum slope and have
relatively little impact on the emission/absorption line strengths.
For this reason any given galaxy spectrum which is projected into the
plane defined by ($pc_1$,$pc_2$) will not be uniquely defined in the
direction of varying continuum but will be robust in the orthogonal
direction (which measures the average line strength).

This robust axis is simply that shown in Fig.1(b) and we denote 
the projection onto this eigenspectrum by $\eta$,
\begin{equation}
 \eta = a\cdot pc_1 + pc_2 \;.
\end{equation}
Where $a$ is a constant which we find empirically to be $a=0.5\pm
0.1$.  

\begin{figure}
 \plottwo{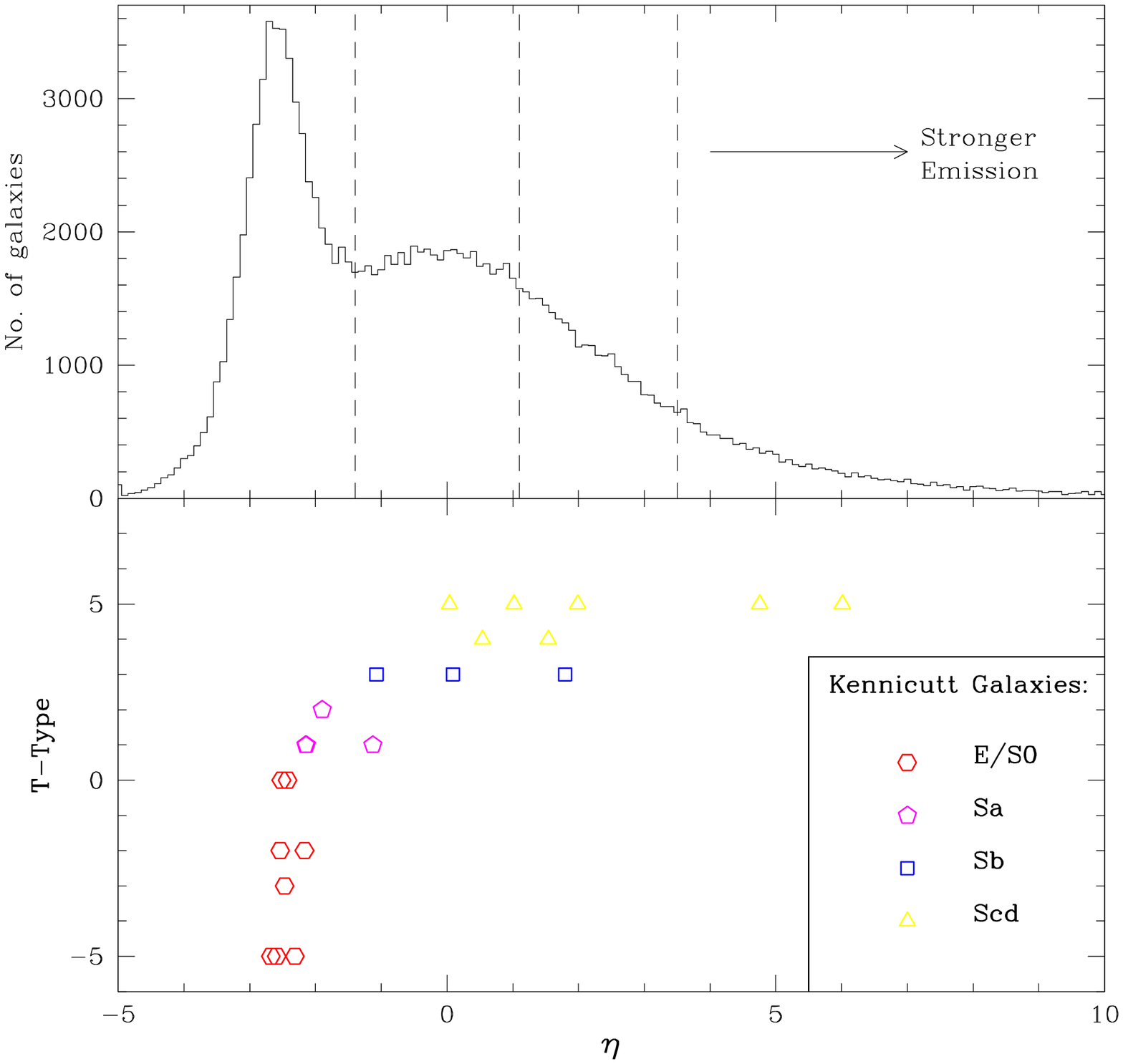}{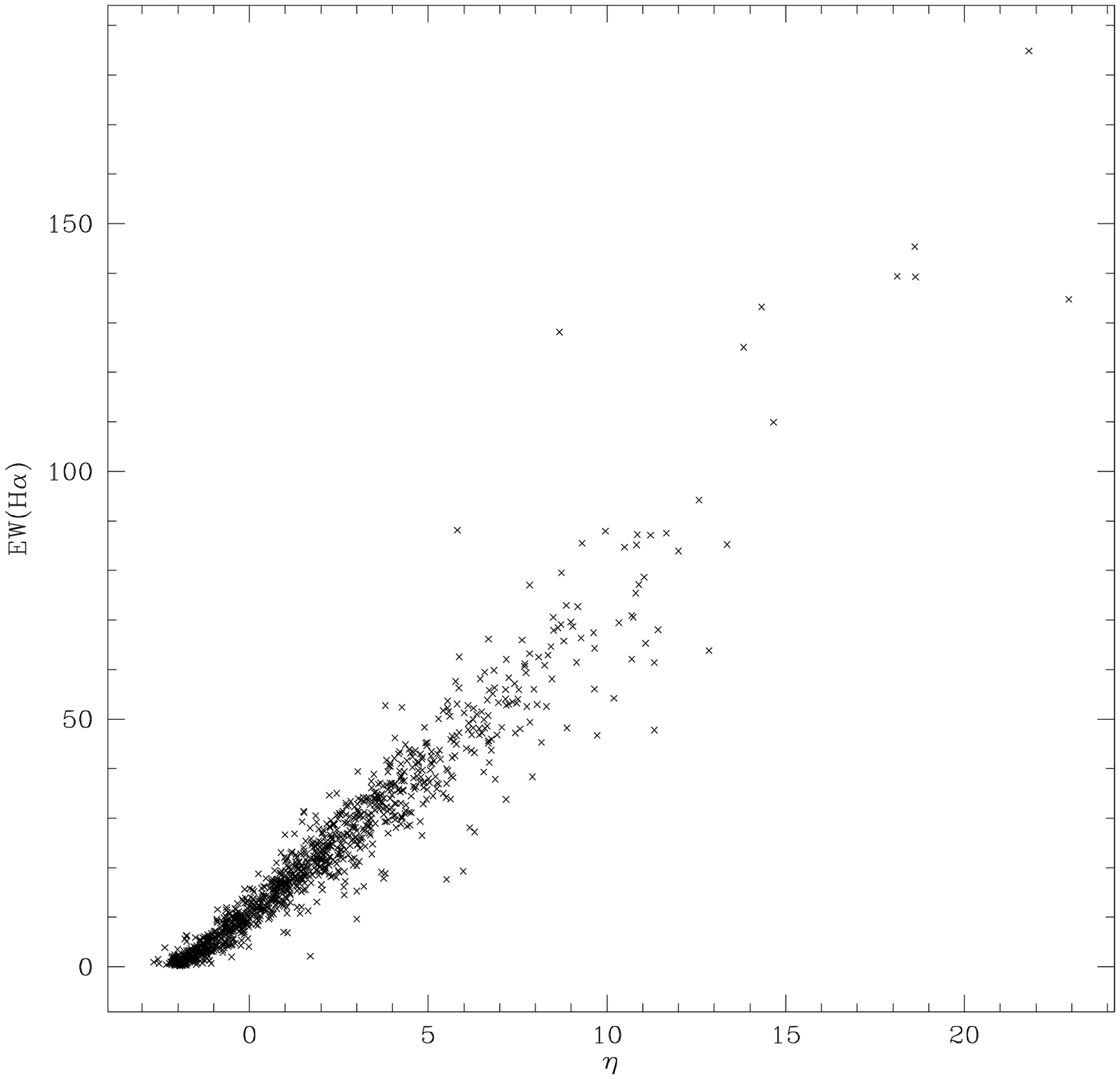}
 \caption{The distribution of $\eta$ projections in the observed 2dF
 galaxies (left).  Also shown in the same plot are the $\eta$ 
projections
 calculated for a subset of the Kennicutt Atlas galaxies.  The second
 plot shows the strong correlation between $\eta$ and EW(H$\alpha$) 
(right).}
 \label{eta}
\end{figure}

The result of this analysis is that we have now identified the
single most (statistically) important component of the galaxy
spectra which is robust to the known instrumental uncertainties,
$\eta$.  We have therefore chosen to adopt this (continuous) variable as 
our measure
of spectral type.  We show the observed distribution of $\eta$
projections for the 2dF galaxies in Fig.~\ref{eta}, together with the 
projections calculated from the Kennicutt Atlas~\cite{kennicutt} galaxies
which have known morphologies.  It can be
seen from this figure that a clear trend exists between $\eta$ and
galaxy morphologies.

Qualitatively,  $\eta$ is a measure of the relative absorption/emission
line strength present in a given galaxies spectrum, and hence can be
interpreted as a measure of the relative current star-formation in
that galaxy.  To highlight this relationship we show in
Fig.~\ref{eta} the strong correlation between the EW(H$\alpha$) and
$\eta$.

It is interesting to note that the distribution displays
some degree of bimodality similar to that observed in the Sloan colour 
distribution~\cite{sloan}.

\section{Applications}

We show below three applications of $\eta$. They illustrate how
$\eta$, which relates to atomic processes,
is connected to the global statistic of galaxies and their
clustering on the $\sim 10 h^{-1}$ Mpc scale.

\subsection{Population Mix in 2dF vs. 2MASS}

\begin{figure}
 \plotone{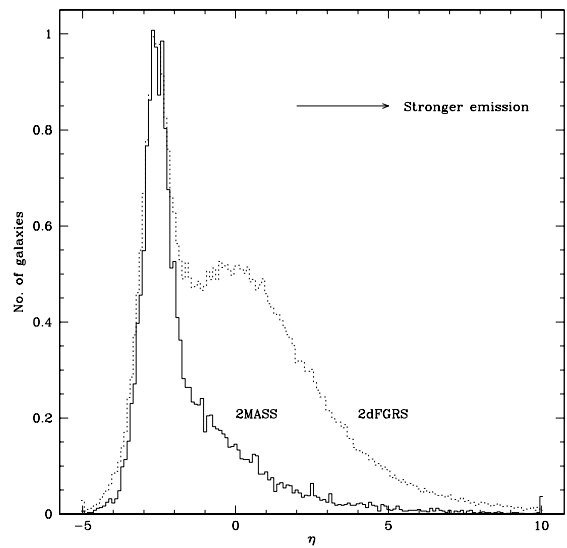}{8cm}
 \caption{The distribution of the spectral type parameter, $\eta$,
 in the full 2dFGRS and the matched 2MASS catalogue~\cite{cole}.
  We see that about two thirds of the 
  2MASS sample are early-type galaxies.}
 \label{2massfig}
\end{figure}

Having a continuous measure of the spectral type of a galaxy allows us
to easily compare the population mix of different
samples of galaxies.  As an example of this we compare the $\eta$ distribution
of the 2dFGRS with that of the overlapping 2MASS galaxy
sample~\cite{cole} ($J_{Kron}<14.45$) in Fig.~\ref{2massfig}.  It is
clear from this figure that the two populations have quite different
compositions.  The Near-IR selected 2MASS galaxies are dominated by
early-type galaxies (the peak roughly corresponds to galaxies with
E/S0 morphologies) which make up approximately two thirds of the
sample.  This contrasts with the $b_J$ selected 2dFGRS galaxies which
contain a much larger selection of star-forming galaxies.

Such comparisons between samples of galaxies  are crucial 
in order to identify the different selection biases in galaxy
surveys, and hence ensure that comparisons of subsequent results
(e.g. luminosity functions) are fair.

\subsection{Luminosity Functions per Spectral Type}

Because $\eta$ is a continuous measure of the spectral type we are now in a 
position to develop bivariate formalisms for such applications as the 
luminosity function $\phi(M,\eta)$.  However, in order to simplify our 
analysis we have instead chosen to divide the distribution in $\eta$ into four
separate types, as shown in Fig.~\ref{eta}.  

\begin{figure}
 \plotone{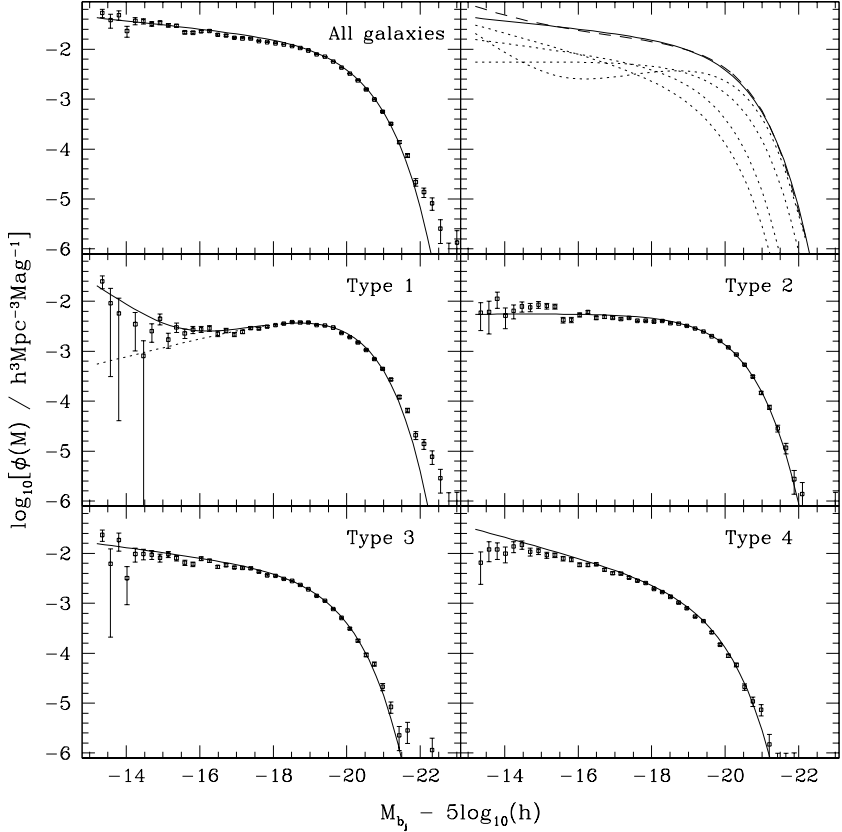}{12cm}
 \caption{The LF derived using the entire galaxy sample is shown in the top
          left plot.  The top right plot shows how this relates to the
          summation of the individual LFs for each spectral type, which are
          shown in the remaining panels. Types 1 to 4 correspond
           to a sequence in $\eta$ from 
           passively to actively star-forming galaxies.}
 \label{lfplot}
\end{figure}

Having defined our 4 spectral types we can now proceed to calculate the
luminosity function (LF) for each of these types using the standard formalisms
of step-wise
maximum likelihood~\cite{efstathiou} and the maximum likelihood fit of a 
Schechter function~\cite{sty} (more detailed descriptions of this analysis have
been given in previous work~\cite{madgwick}).  These LFs are plotted in 
Fig.~\ref{lfplot}.

A general trend is clear in that galaxies with relatively more current
star-formation systematically have a fainter characteristic magnitude and a
steeper faint-end slope.  It is also interesting to note that whilst
the Schechter function provides a good fit to the LFs of the late-type
galaxies it does not fit the faint early-type galaxies well, perhaps 
suggesting the presence of a significant dwarf population.  This
feature is not evident in the overall LF and so would not have been
found without having first performed this classification.

\subsection{Clustering per Spectral Type}

\begin{figure}
 \plottwo{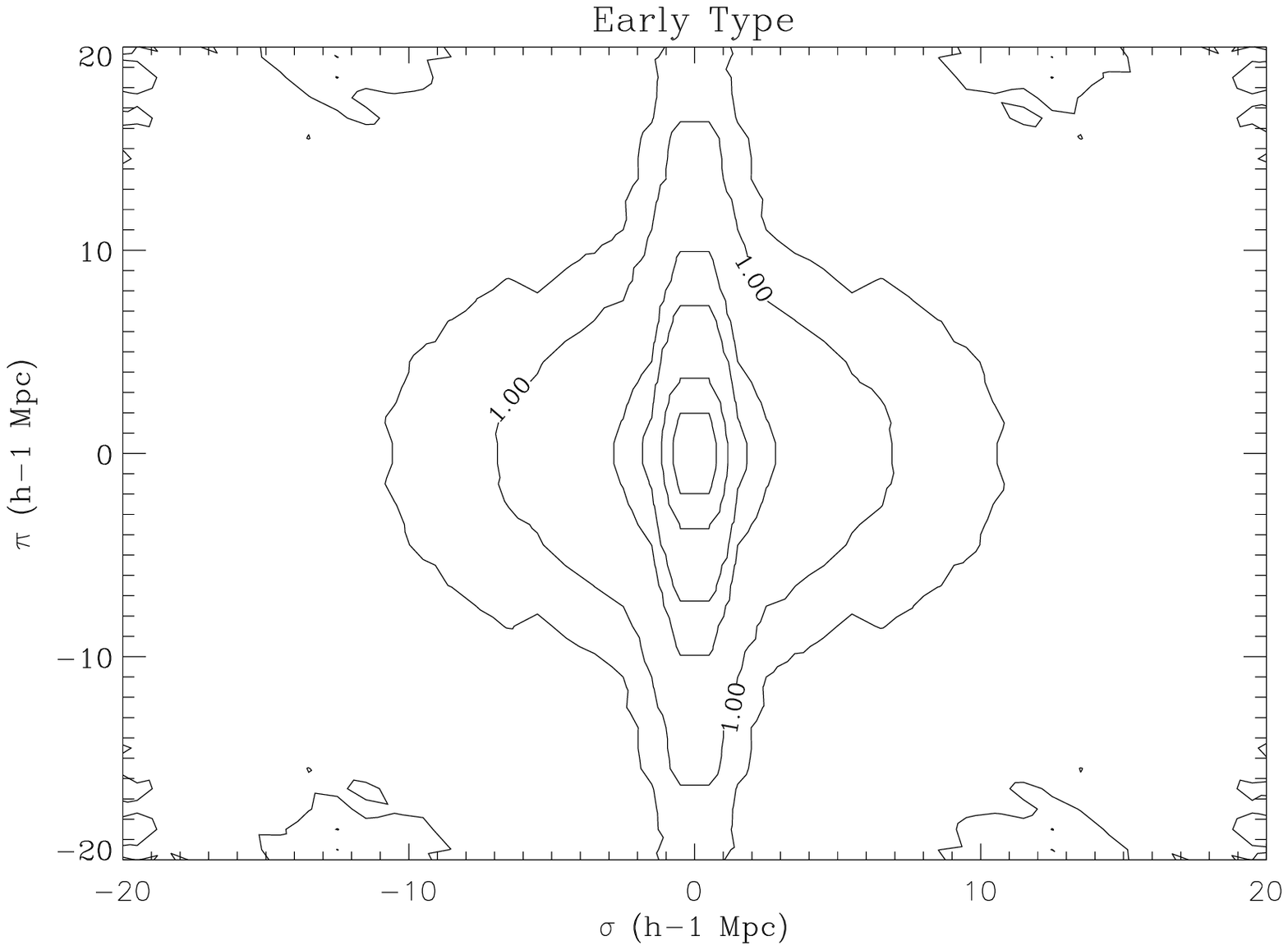}{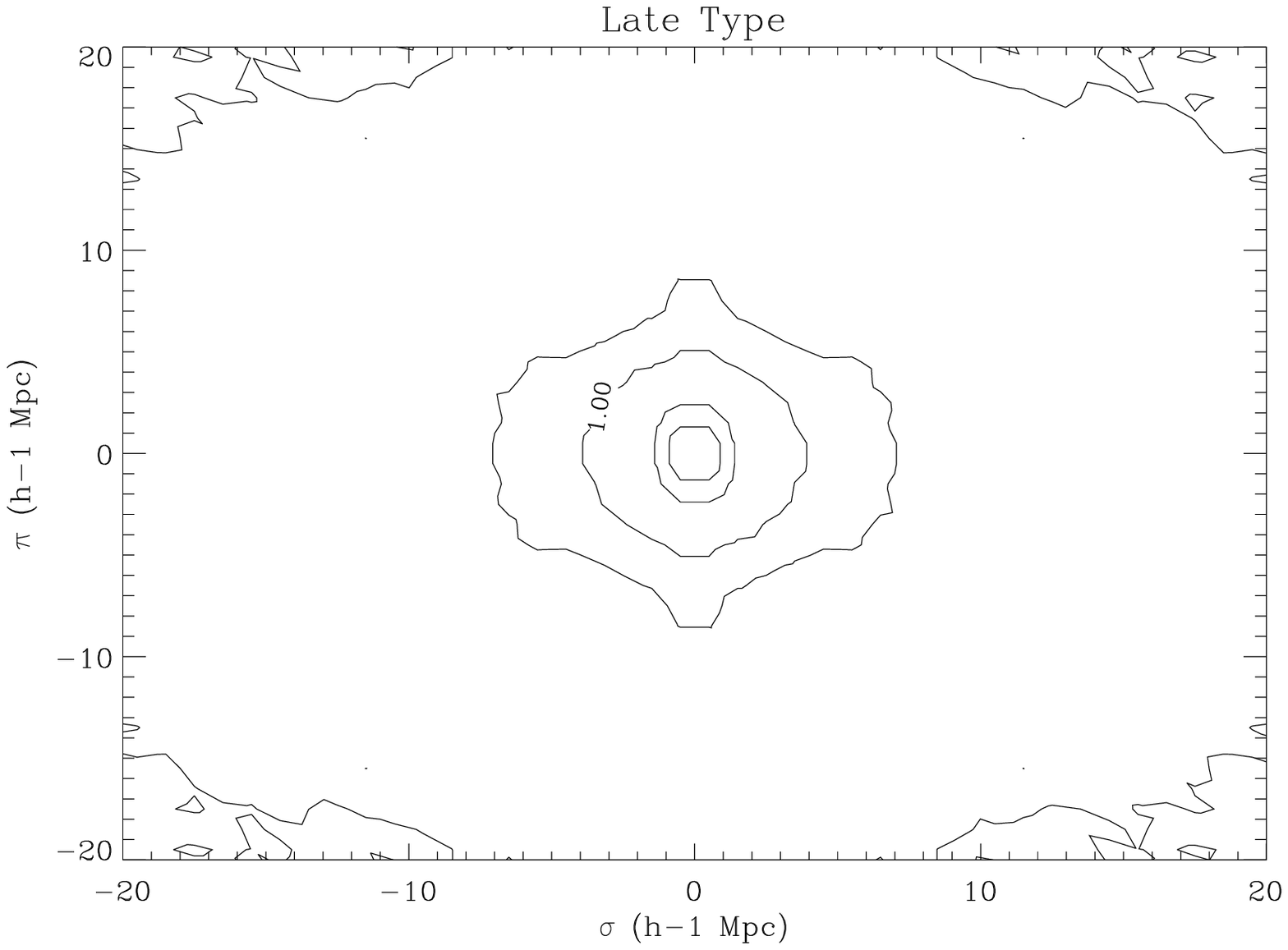}
 \caption{The two point correlation function $\xi(\sigma,\pi)$ plotted for
   passively (left) and actively (right) star-forming galaxies 
   (right).  The contour levels are 15,10,5,3,1 (labelled),0.5,0.}
 \label{xiplot}
\end{figure}

In order to investigate the different clustering properties of different types 
of galaxies we make use of the two point correlation function~\cite{norberg}.
%Because the clustering of galaxies varies with the absolute 
%magnitude of
%the galaxies being sampled~\cite{norberg} we restrict the range in absolute 
%magnitude
%considered to a narrow range around $M^*$.  
The calculated correlation 
function is shown in terms of the line-of-sight 
and perpendicular to the line-of-sight separation $\xi(\sigma,\pi)$ 
in Fig.~\ref{xiplot}.  In this plot we show the correlation function calculated
from the most passively (type 1) and actively (types 3 and 4) star-forming 
galaxies.

%\vspace{12cm}

It can be seen from Fig.~\ref{xiplot} that the clustering properties of the
two samples are quite distinct.  The more passively star-forming (`red')
galaxies 
display a prominent `finger of God' effect and also have a higher overall
normalisation than the more actively star-forming (`blue') galaxies.
This is a manifestation of the well-known morphology-density relation.
This diagram allows us to determine the combination of the mass density 
and biasing parameter $\beta = \Omega_m^{0.6}/b$ (\cite{peacock}) 
per spectral type (\cite{norberg2}, \cite{hawkins}) 

\section{Conclusions}

The new generation of extremely large astronomical surveys currently 
taking place is ushering in a new
era of precision astronomy.  However, in order to reduce the vast quantities
of information being produced into more practical data sets one must 
first address 
the challenges of compressing the data in a meaningful way.  

As an example
of this we have demonstrated here in these proceedings how the large number of 
galaxy spectra being observed in the 2dFGRS can be reduced in the regime of
a simple spectral classification, $\eta$.  We have addressed the issues of
ensuring that this classification is meaningful and also robust to the known
instrumental uncertainties.  In addition we have shown how this classification
can be applied to the calculation of luminosity and correlation functions in 
a way which adds new constraints on scenarios for galaxy formation and 
biasing. 

\section{Acknowledgements}

We thank our 2dFGRS collaborators\footnote{The 2dFGRS Team comprises:
      I.J.Baldry, C.M.Baugh, J.Bland-Hawthorn, T.J.Bridges, R.D.Cannon, S.Cole,
      C.A.Collins,  
      M.Colless,
      W.J.Couch, N.G.J.Cross, G.B.Dalton, R.DePropris, S.P.Driver,
      G.Efstathiou, R.S.Ellis, C.S.Frenk, K.Glazebrook, E.Hawkins, C.A.Jackson,
      O.Lahav, I.J.Lewis, S.L.Lumsden, S.Maddox, 
      D.S.Madgwick, S.Moody, P.Norberg, J.A.Peacock, B.A.Peterson,
      W.Sutherland, K.Taylor }
for their contribution to
the work presented here.

\vfill
\end{document}